\documentclass[11pt,a4paper]{article}
\pdfoutput=1
\usepackage{jinstpub}
\usepackage{color}
\usepackage{lineno,hyperref,xspace,topcapt}
\modulolinenumbers[5]

\title{Brightness and uniformity measurements of plastic scintillator
  tiles at the CERN H2 test beam}

\collaboration{CMS HCAL Collaboration}

\emailAdd{abelloni@umd.edu}

\newcommand{\nm}{\ensuremath{\,\mathrm{nm}}\xspace}
\newcommand{\mm}{\ensuremath{\,\mathrm{mm}}\xspace}
\newcommand{\ns}{\ensuremath{\,\mathrm{ns}}\xspace}
\newcommand{\s}{\ensuremath{\,\mathrm{s}}\xspace}
\newcommand{\fC}{\ensuremath{\,\mathrm{fC}}\xspace}
\newcommand{\GeV}{\ensuremath{\,\mathrm{Ge\hspace{-.08em}V}}\xspace}
\newcommand{\MeV}{\ensuremath{\,\mathrm{Me\hspace{-.08em}V}}\xspace}

\abstract{
  We study the light output, light collection efficiency and signal
  timing of a variety of organic scintillators that are being
  considered for the upgrade of the hadronic calorimeter of the CMS
  detector. The experimental data are collected at the H2 test-beam
  area at CERN, using a 150\GeV muon beam. In particular, we
  investigate the usage of over-doped and green-emitting plastic
  scintillator, two solutions that have not been extensively
  considered.  We present a study of the energy distribution in
  plastic-scintillator tiles, the hit efficiency as a function of the
  hit position, and a study of the signal timing for blue and green
  scintillators.
}


\begin{document}

\maketitle

\flushbottom

\section{Introduction}
\label{sec:introduction}

The analysis of the collision and calibration data collected by the
CMS Collaboration at the end of Run-I (fall 2012) demonstrated that
the light yield of the hadronic endcap calorimeter is decreasing due
to radiation damage much faster than anticipated
(e.g.,~\cite{Contardo:2020886}, Ch.~3). In particular, it is estimated
that the hadronic calorimeter will not be able to survive until the
end of its originally planned lifetime without the replacement of part
of its active material. Moreover, it is estimated that the material
currently used is not radiation-tolerant enough to be usable for a
replacement detector, since it would degrade too fast in the time
interval between scheduled upgrades of the CMS detector.

We set to investigate the performance of a few plastic scintillators
that seem to offer an increased radiation tolerance with respect to
SCSN-81, the scintillator currently used in the CMS hadronic
calorimeter. The improved radiation tolerance is obtained in two
different ways: increasing the concentration of the scintillating
dopants; using green-emitting dopants instead of the more common
blue-emitting ones. Increasing the dopant concentration has the effect
of reducing the light yield of an unirradiated scintillator because
light self-absorption by the dopant is increased. However, as
radiation damages the scintillator base, and reduces its attenuation
length, increasing the dopant concentration enhances the probability
that light is wavelength-shifted by the dopant, and then more
efficiently travels through the damaged base. A similar effect is
obtained by using green-emitting dopants, in which the scintillation
light is emitted at longer wavelengths, thus being less sensitive to
damage in the scintillator base.

Experimental data are collected at the H2 test-beam area at CERN,
where an asynchronous 150\GeV muon beam is available. We study the
energy response, measured in units of integrated charge, of different
scintillator tiles when they are traversed by high-energy muons, the
detection efficiency as a function of the muon position on the tile,
the integrated charge distribution in 25\ns time slices,
and the distribution of the difference between the signal times of
pairs of scintillator tiles. The last measurements are of particular
interest for green-emitting scintillators, which are usually
considered to be considerably slower than blue-emitting scintillators.

\section{Experimental Area}
\label{sec:experimental_area}

This experiment was conducted at the H2 Beam Line located at the North
Area of the CERN accelerator complex, in the Prevessin site. The H2
Beam Line is serviced by the Super Proton Synchrotron (SPS). The SPS
is capable of accelerating protons to 450\GeV; protons are then
extracted from the SPS accelerator, and directed towards the T2
target. The emerging beam is filtered to remove secondary particles
such as electrons, and the momentum of the remaining muons is
selected to be 150\GeV. 

\begin{figure}[!ht]
  \begin{center}
    \includegraphics[trim={0 6cm 0 5cm},clip,width=0.75\textwidth]{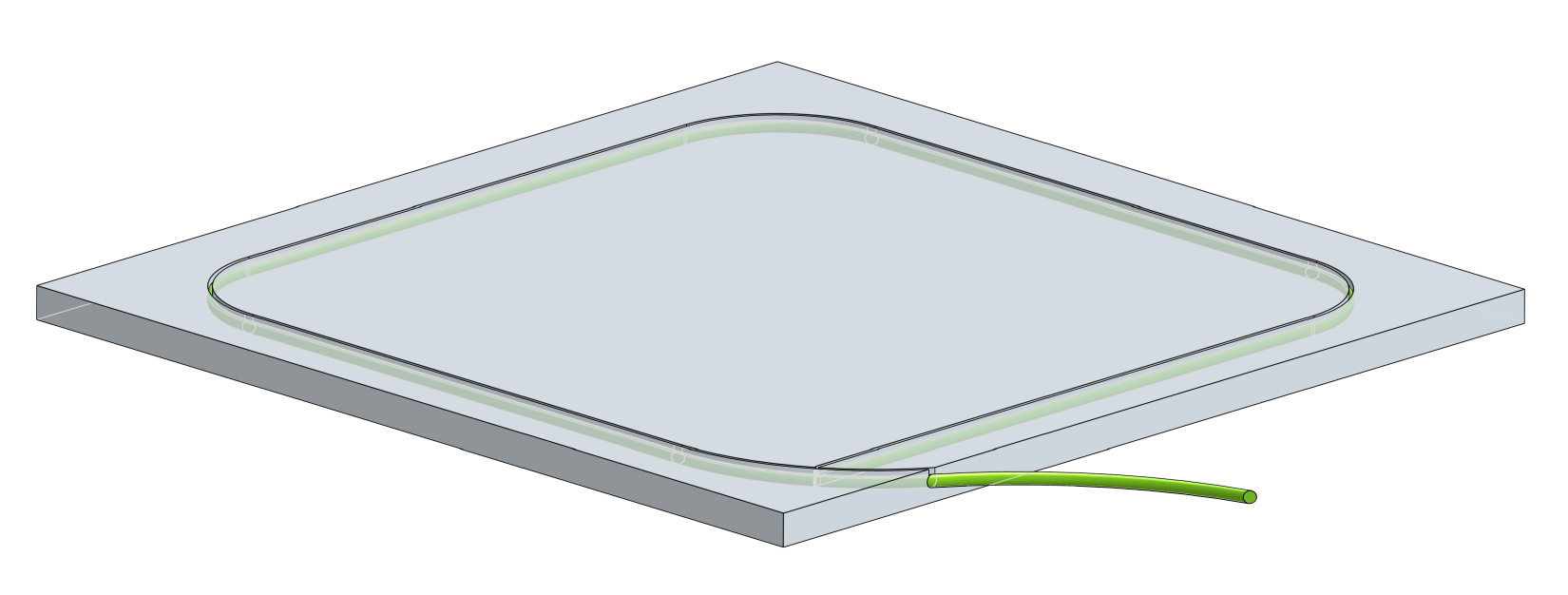}
    \caption{Drawing of a scintillator tile, including the WLS
      fiber installed inside a $\sigma$-shaped groove.}
    \label{fig:tile}
  \end{center}
\end{figure}

The devices under test are $100\times100\times4\mm^3$ plastic
scintillator tiles. A $\sigma$-shaped groove is carved into each tile;
a wavelength-shifting fiber (WLS) is installed inside the grooves, as shown
in figure~\ref{fig:tile}. The plastic tiles are wrapped in
Tyvek\footnote{Tyvek B1060 (registered trademark of DuPont Co.) is a
thin sheet of high density white polyethylene.} paper, and enclosed
in black-plastic 3D-printed boxes, which offer two distinct services:
they provide a support for the tile, thus reducing the risk of
breaking a readout fiber by accident, and simplify keeping the tiles
in a dark environment, thus reducing noise. The Tyvek wrapping has
been demonstrated to enhance the light yield by a factor of two, by
diffusing the light that reaches the scintillator walls back into the
tile. The tiles are located on top of a moving table. The table is
aligned in such a way that, as best as possible, the muon beam is
centered in the middle of the scintillator tiles.

Prior to reaching the table, the beam path passes through several
instruments for triggering and precise tracking. In order of
increasing distance from the beam extraction point, the following
instruments are installed: wire chamber "A", wire chamber "B", four
large scintillating plastic trigger counters, and wire chamber "C".

A trigger signal is produced by requiring a coincidence among selected
trigger counters. The first and fourth trigger counters are
$140\times140\mm^2$, the second counter is $40\times40\mm^2$, and the
third counter is $20\times20\mm^2$. In this analysis, the coincidence
between the first and fourth trigger counters was used to select
events. While this trigger selection is larger than the actual size of
the tiles, the usage of tracking information from the wire chambers
allows us to define during the analysis whether a muon had crossed a
tile or not.

\begin{figure}[!ht]
  \begin{center}
    \includegraphics[width=0.495\textwidth]{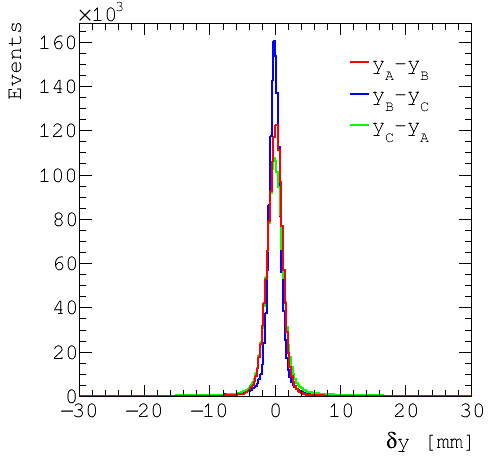}%
    \includegraphics[width=0.495\textwidth]{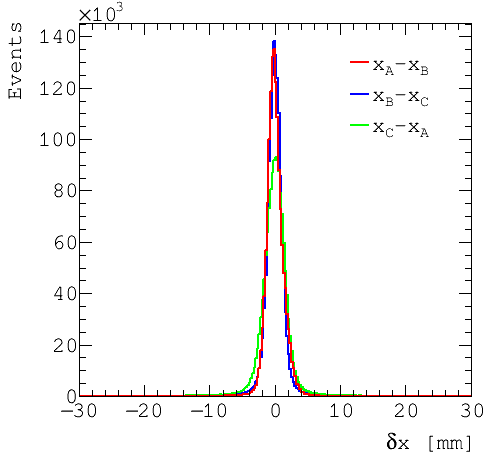}
    \caption{Hit position difference along the $y$ (left) and $x$
      (right) directions between each pair of wire chambers. The
      distributions contain an offset correction calculated assuming
      that muons are travelling along a straight line.}
    \label{fig:deltaxy}
  \end{center}
\end{figure}

The three wire chambers are $100\times100\mm^2$ drift chambers, each
of which has a resolution of about 0.5\mm in the $x$ and $y$
directions, perpendicular to the direction of the beam. The position
measurements of the three chambers are used to determine the
trajectory of each muon and hence its intersection point with each
tile. An alignment procedure is established by assuming that the muons
are moving along straight lines, and therefore that the distribution
of position differences between pairs of wire chambers must be
zero-mean Gaussian-distributed. Figure~\ref{fig:deltaxy} shows the
distribution of the position differences along the $y$ and $x$
directions for each pair of wire chambers, after the alignment
corrections.

The data acquisition system is built with the front-end and back-end
electronics designed for the Phase-I upgrade of the CMS hadronic
calorimeter (HCAL). The system is described in detail
in~\cite{Mans:1481837}. The light produced by the scintillator is
transmitted via a wavelength-shifting fiber, connected to a plastic
clear fiber, to a silicon photomultiplier (SiPM). The current pulse
produced by the SiPM is integrated by a charge-integration-and-encoder
(QIE;\cite{Mans:1481837}, Ch.3.1) chip, expressively designed for the
CMS HCAL detector. The encoded signal is transmitted, via an optical
link, to the back-end electronics; encoded signals from the QIE chips
and the wire chambers are then built into events, which are saved to
disk. The beam is asynchronous, in that muons are produced without a
fixed timing structure in spills containing about thirty thousand
muons within a 10\s window. About two spills per minute are provided.

\section{Scintillators}
\label{sec:scintillators}

The scintillators under test are cut in $100\times100\times4\mm^3$
tiles, with a $\sigma$-shaped groove carved in the plastic, which
holds a wavelength-shifting fibers. The materials at disposal include
both commercial scintillators and specially-formulated ones. A sample
of the scintillator used in the construction of the current CMS HCAL
system is also used, from the same production lot. This scintillator,
SCSN-81, is not produced any more, but commercial equivalents are
available. Table~\ref{tab:scintillators} summarizes the main features
of the scintillators tested. The other scintillators under test are
produced by Eljen Technology.~\cite{Eljen} Studies of how radiation
affects the transmission of light through some of these plastic
scintillators have been presented in~\cite{Liao:2015zsa}.

\begin{table}[htp]
\begin{center}
\topcaption{\label{tab:scintillators}
  Main characteristics of scintillators included in analysis.}
\begin{tabular}{lccp{0.45\linewidth}}
\hline\hline
Material & Base & Emission Peak & Notes \\
\hline
SCSN-81 & PS & 440\nm & Material used in CMS HCAL detector,
originally produced by Kuraray, which later stopped its
production\\

EJ-200 & PVT & 425\nm & Commercial scintillator, produced by Eljen
Technologies. Equivalent to St.Gobain BC-408\\

EJ-200 2X & PVT & 425\nm & Special version of EJ-200 specifically
produced by Eljen Technologies. The concentration of primary dopant is
doubled with respect to the commercial version of EJ-200\\

EJ-200 P2 & PVT & 425\nm & Special version of EJ-200 specifically
produced by Eljen Technologies. A different type of primary dopant is
used\\

EJ-260 & PVT & 490\nm & Green-emitting scintillator. Its decay time
is 9.2\ns, while the typical decay time of blue scintillators is
about 2\ns\\
\hline\hline
\end{tabular}
\end{center}
\end{table}

The majority of the tiles are fabricated from blue-emitting
scintillator. The wavelength-shifting fiber that matches the peak of
their emission is Y11, produced by Kuraray. The re-emitted light peaks
at 476\nm, in the green range. EJ-260 is instead a green-emitting
scintillator. It is matched to a O2 wavelength-shifting fiber, also
produced by Kuraray. The emission peak of the O2 fiber is 538\nm, in
the red/orange range. The emission spectra of the Y11 and O2
wavelength-shifting fibers are reported in~\cite{Kuraray}, while the
emission spectra of the EJ-200 and EJ-260 scintillators are reported
in~\cite{Eljen}, respectively.\footnote{The production of SCSN-81 has
  been discontinued, and no documentation about this scintillator is
  available on the Kuraray website. A copy of an old Kuraray catalog
  containing SCSN-81 spectra is available at~\url{http://www.phenix.bnl.gov/WWW/publish/donlynch/RXNP/Safety\%20Review\%206_22_06/Kuraray-PSF-Y11.pdf}.}

\begin{figure}[!ht]
  \begin{center}
    \includegraphics[width=0.75\textwidth]{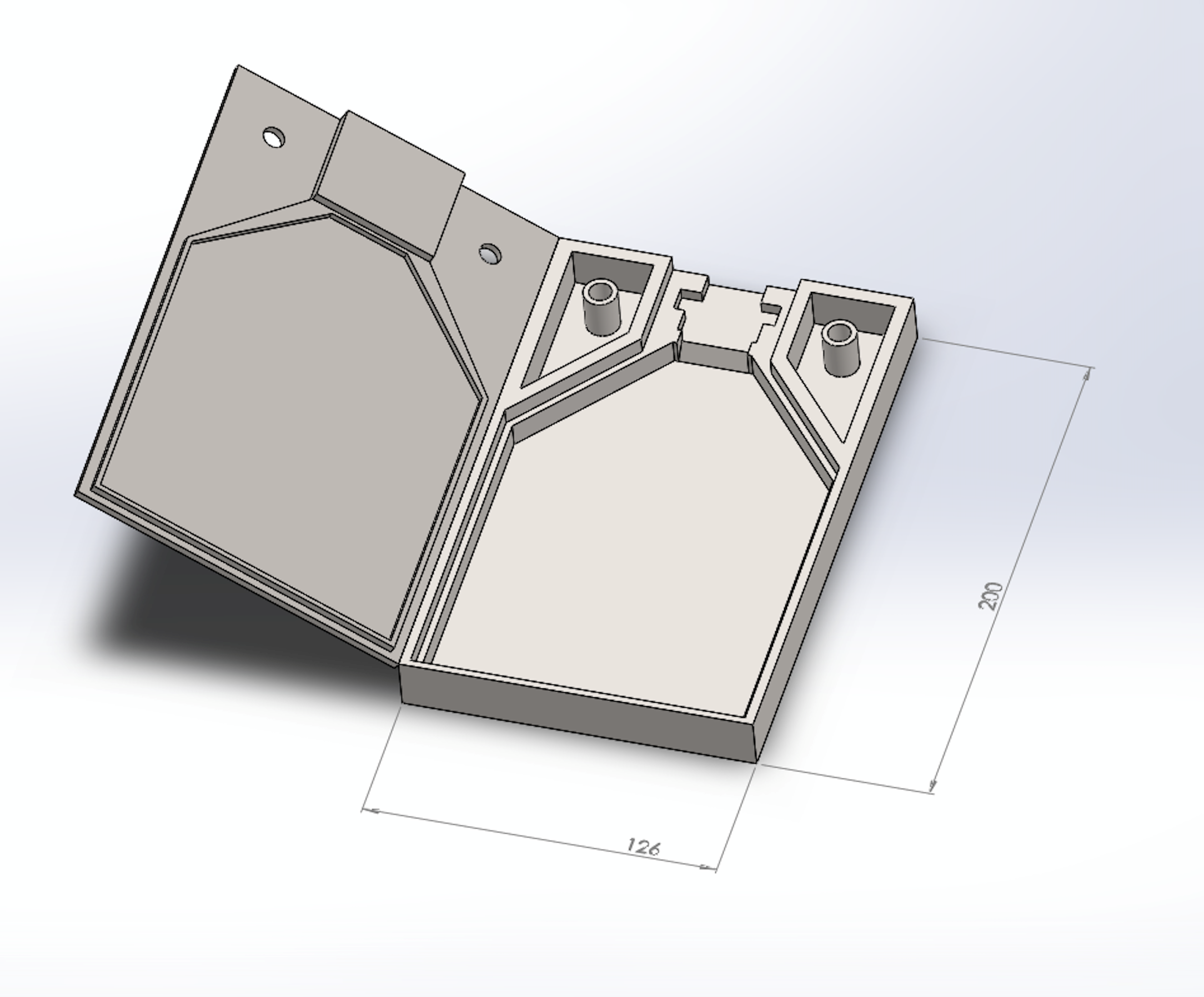}
    \caption{Drawing of a container for plastic tiles; its dimensions
      are expressed in millimeters.}
    \label{fig:box}
  \end{center}
\end{figure}

Each tile is individually housed inside a 3D-printed black plastic
box. An example of this installation is shown in figure~\ref{fig:box}. A
scintillator tile is set inside the box, and its wavelength-shifting
fiber is connected to a 12-fiber connector firmly blocked by plastic
supports. The Tyvek wrapping has been removed to show the $\sigma$
pattern of the wavelength-shifting readout fiber.  This setup greatly
reduces the probability of breaking the wavelength-shifting fiber by
inadvertently pulling or twisting the connector.

\section{Data Samples}
\label{sec:data_samples}

The data sample before any selection is about 5 million triggered
events. Events are required to have a single muon hitting each of the
three wire chambers. This ensures that the energy measurement
correspond to a single MIP signal, and allows us to map the detection
efficiency as a function of the hit position on a tile.  This
requirement selects a data sample of 1,416,624 events.

\section{Charge Analysis}
\label{sec:charge_analysis}

\begin{figure}[!ht]
  \begin{center}
    \includegraphics[width=0.75\textwidth]{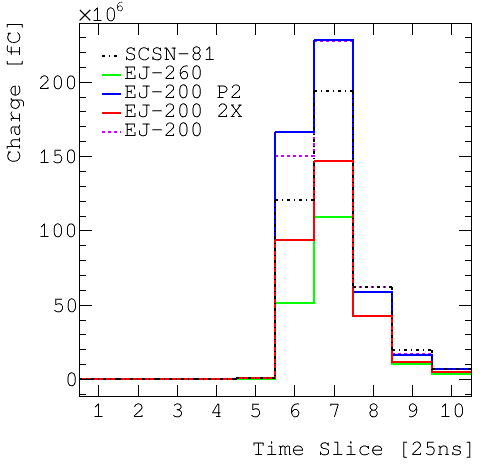}
    \caption{Distribution of integrated charge per 25\ns time slice in
      the whole data sample.}
    \label{fig:ts}
  \end{center}
\end{figure}

The energy loss of minimum-ionizing particles crossing a scintillator
tile is reported in units of integrated charge. The DAQ system reports
the energy measured in time slices corresponding to 25\ns
each. Figure~\ref{fig:ts} presents the energy distribution per time
slice, which shows that the energy corresponding to a hit is
distributed in time slices 6 to 9. It also shows that all the channels
corresponding to different scintillators are timed-in similarly; this
allows us to select the same time slices for all scintillators. The
plot also contains an indication that the EJ-260/O2
scintillator/WLS-fiber system is slightly slower than the others: its
corresponding energy distribution is spread more among time
slices. The fraction of integrated charge in the first two time slices
(6-7) is 81\% for the blue EJ-200 scintillator, and 74\% for the green
EJ-260 scintillator, with negligible uncertainties.

A more precise analysis of the arrival time structure of signals in
different scintillators is presented in section~\ref{sec:time}.

Integrated charge spectra report the following features:

\begin{itemize}
  \item a pedestal peak, corresponding to signal observed when no
    particle hits the tile: electronic noise; SiPM dark current; light
    leaking into the tile. The pedestal peak is typically located
    below 25\fC.
  \item a set of peaks extending up to a few thousand of
    femtocoulombs. These correspond to photo-electrons produced by the
    passage of a muon through the tile. It is interesting to note that
    the energy resolution of a SiPM allows us to distinguish each peak
    corresponding to a different number of photo-electrons.
\end{itemize}

We require that the integrated charge be larger than 25\fC to identify
MIP hits. This requirement is set by observing an energy spectrum
without any selection cut on the muon hit positions, i.e., without
enforcing the passage of a muon through the tile.

Second, it is necessary to identify the position of the tiles with
respect to the beam and the wire chambers. We perform a
2-dimensional measurement of the hit efficiency as a function of $x$
and $y$. The hit efficiency is defined as the fraction of events in an
$x$-$y$ cell with an integrated charge larger than 25\fC.

\begin{figure}[!ht]
  \begin{center}
    \includegraphics[width=0.75\textwidth]{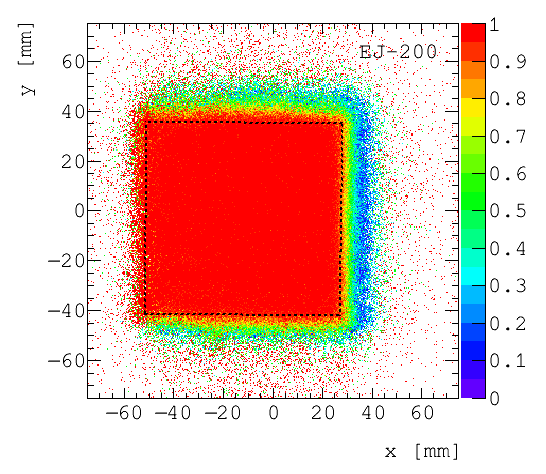}
    \caption{Map of the hit efficiency as a function of the hit
      position in the EJ-200 tile. Scintillator tiles are not
      completely centered with respect to wire-chamber position; the
      areas where the efficiency is about $20\%$ correspond to a
      location not covered by plastic scintillator. The efficiency is
      not zero because a small signal due to electronic noise or light
      leaks can still be recorded. The dashed lines identify a
      fiducial area within the scintillator tile. The
      light-collection efficiency of a tile is measured using
      exclusively muons crossing that fiducial
      area.}
    \label{fig:effmaps}
  \end{center}
\end{figure}

The efficiency map for the EJ-200 tile is shown in figure~\ref{fig:effmaps}; 
similar maps have been produced for the other tiles under test, and show
the same features. It is clear from the map that as soon as a
tile is centered, its hit efficiency exceeds $95\%$, and that
the efficiency is rather uniform on the tile. This is a result of the
$\sigma$ shape adopted for the wavelength-shifting fiber. Its shape
has been optimized to enhance uniformity of light-collection
efficiency.~\cite{deBarbaro:1991yn}

\begin{figure}[!ht]
  \begin{center}
    \includegraphics[width=0.495\textwidth]{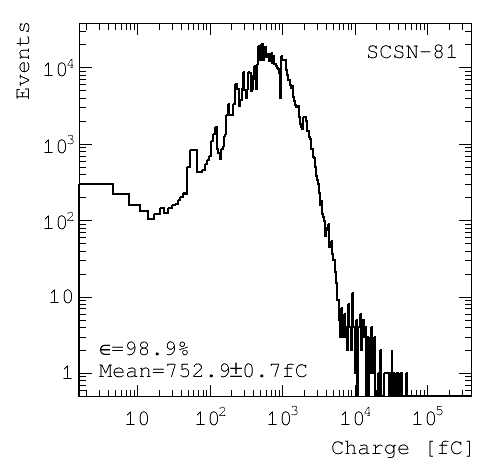}%
    \includegraphics[width=0.495\textwidth]{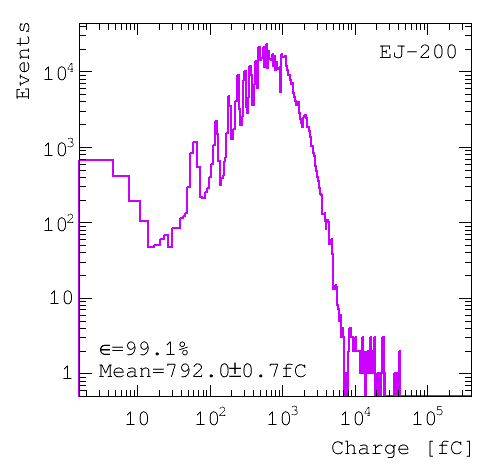}
    \includegraphics[width=0.495\textwidth]{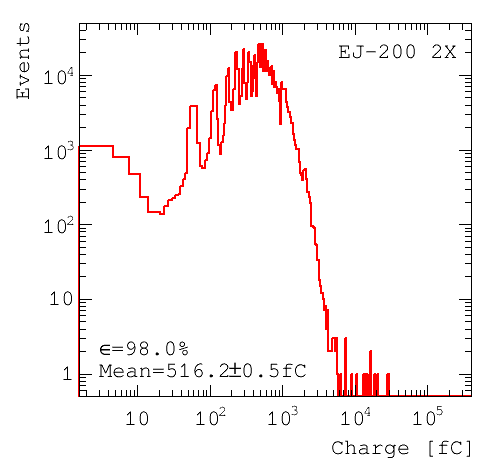}%
    \includegraphics[width=0.495\textwidth]{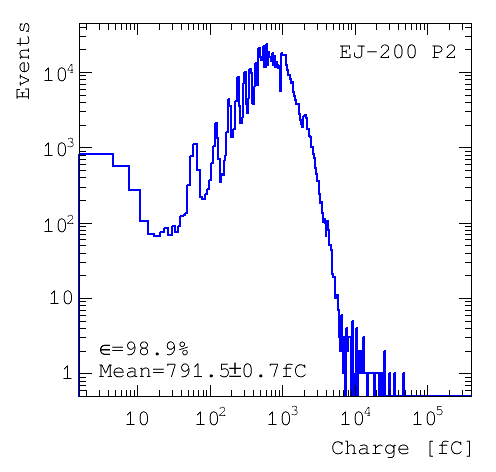}
    \includegraphics[width=0.495\textwidth]{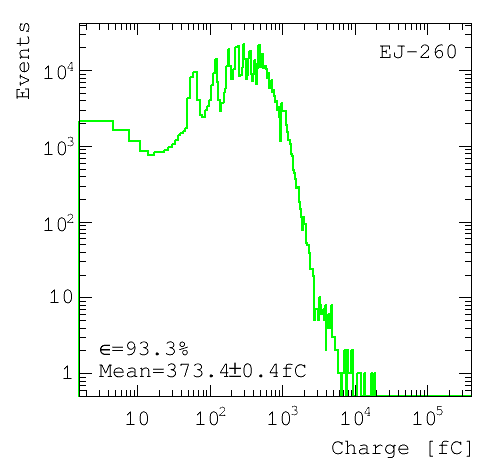}
    \caption{Integrated charge per muon in femtocoulombs. Events are
      selected requiring that a single muon be crossing the
      scintillator tile. The fraction of events in the plot with
      integrated charge above 25\fC represents the hit
      efficiency.}  \label{fig:fidenergy} \end{center}
\end{figure}

The efficiency maps allow us to determine the position of each
scintillator tile with respect to the wire-chamber measurements. It is
therefore possible, after inspecting the efficiency maps, to set a
tile-dependent $x$ and $y$ cut that selects only events in which a
single muon crosses the active material of a tile. These requirements
allow us to define the light-collection efficiency, i.e., the
probability that a muon crossing a tile produces a signal above
25\fC. The distributions of the energy released by muon MIP crossing
the fiducial area within a tile is shown in figure~\ref{fig:fidenergy}.

We note that the over-doped sample of EJ-200 has a smaller average
number of photo-electrons collected per muon MIP with respect to
SCSN-81, which we will use as reference material. This is consistent
with the expectation: the increase in dopant concentration causes a
reduction in light yield by enhancing self-absorption by the dopant
itself. We similarly observe that the light yield of EJ-260 is also
smaller than SCSN-81. This can be partly explained by the fact that
EJ-260 is less bright than EJ-200 (9,200 photons per 1\MeV electron
versus 10,000), the quantum efficiency of SiPM is lower for red/orange
light than for green light~\cite{Heering:2016aqi}, and the
green-to-red/orange conversion in WLS fibers is less efficient than the
blue-to-green one.

\section{Time Analysis}
\label{sec:time}

The upgraded front-end electronics allows for the first time the
precise measurement of the time at which a signal pulse is
produced. The resolution of the time measurement is 0.5\ns, as it has
been checked by injecting light into a tile with a laser, and using
the laser itself to produce a trigger signal. Unfortunately, it is not
possible to produce an absolute timing measurement because there is no
available time reference in the beam data. Muons arrive
asynchronously, during a spill that lasts about 10\s. It is not
possible to measure when a muon hits a tile, and compare that time
with the measurement of the electronics, which would allow one to
measure the decay time of each scintillator.

However, it is possible to compare how fast the scintillators are with
respect to each other by looking at the difference between the signal
timing of two scintillators. This distribution does not depend on the
time at which the muon hits the tiles, because it is the same for all
tiles. This distribution can also be analytically modelled by assuming
that it is the difference between two Gaussian distribution, each of
which corresponds to the distribution of pulse leading-edge times in
one of the scintillators that are being compared.

\begin{figure}[!ht]
  \begin{center}
    \includegraphics[width=0.495\textwidth]{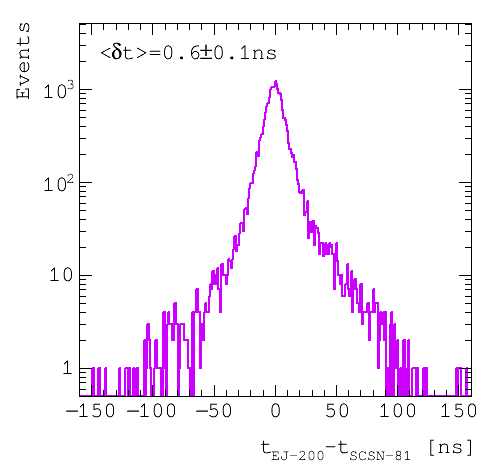}%
    \includegraphics[width=0.495\textwidth]{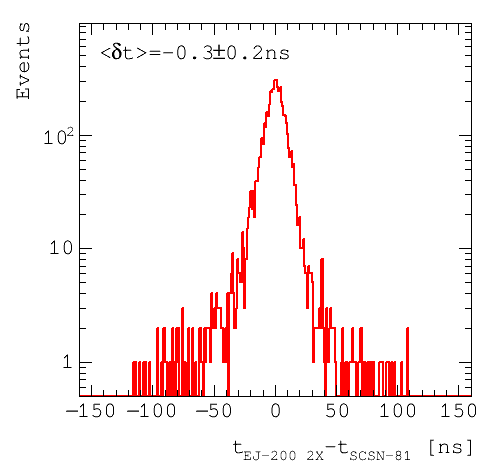}
    \includegraphics[width=0.495\textwidth]{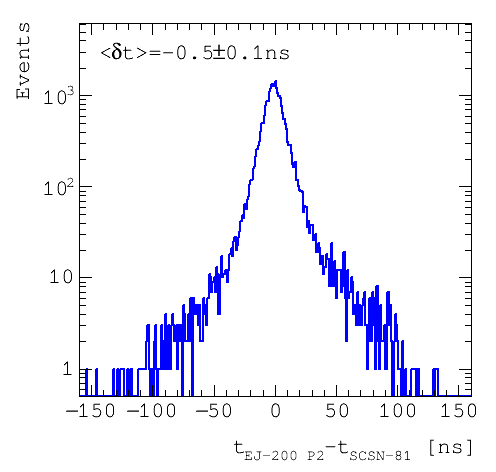}%
    \includegraphics[width=0.495\textwidth]{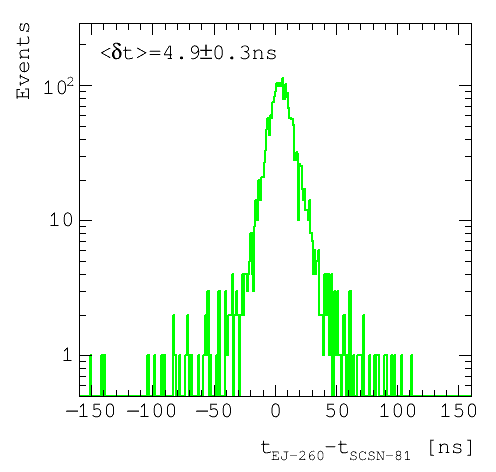}
    \caption{Distribution of the time difference between signals
      obtained in pairs of scintillator tiles. Events are selected
      requiring that a single muon be crossing the scintillator
      tile. The distributions show that blue-emitting scintillators
      are roughly equivalent (the time-difference distributions are
      centered at 0\ns), while the green-emitting EJ-260 scintillator
      is slower (the time-difference distribution with EJ-260 is
      centered at about 5\ns).}
   \label{fig:time}
  \end{center}
\end{figure}

The distributions of time differences are presented in
figure~\ref{fig:time}. They show that the blue scintillators (SCSN-81,
different flavors of EJ-200) have a similar timing; the corresponding
distributions are roughly centered at 0\ns. The EJ-260 green
scintillator is slower than the blue-emitting
scintillators. The mean of the time-difference distribution is about
5\ns. However, we do not think that this precludes the usage of
EJ-260 for a calorimeter at the LHC, where the bunch-crossing
separation is about five times larger, 25\ns. It has also been noted
in section~\ref{sec:charge_analysis} that the EJ-260 signal is
wider in time than the signal from blue scintillators.

\section{Conclusion}

We measure the performance of plastic scintillator tiles using a
150\GeV muon beam at the CERN H2 test-beam area. The materials are
tested for light-collection efficiency and light yield. Commercially
available scintillators are compared to custom-made ones, in an
attempt at investigating the possibility of increasing their radiation
tolerance. Over-doping a scintillator and shifting to a longer
wavelength-emitting dopant are considered two simple methods to
increase the radiation tolerance of a scintillator. The analysis of
light yield and signal timing indicate that both methods are viable,
and do not prevent their application to produce a usable scintillator.

\acknowledgments The authors would like to thank CERN for the
operations of the SPS accelerator; Dragoslav Lazic for supporting
operations in the H2 test-beam area; Janina Gielata (FNAL) for the
preparation of optical connections. This work was supported in part by
U.S. Department of Energy Grants.

\bibliography{DN-16-019_arxiv}

\clearpage
\section*{The CMS HCAL Collaboration}
\addcontentsline{toc}{section}{The CMS HCAL Collaboration}
\parindent 0 pt

\textbf{Yerevan Physics Institute, Yerevan, Armenia}\\
S.~Chatrchyan, A.M.~Sirunyan, A.~Tumasyan\\

\textbf{National Centre for Particle and High Energy Physics, Minsk, Belarus}\\
A.~Litomin, V.~Mossolov, N.~Shumeiko{\def\thefootnote{\fnsymbol{footnote}}\footnotemark[2]}\\

\textbf{Universiteit Antwerpen, Antwerpen, Belgium}\\
M.~Van~De~Klundert, H.~Van Haevermaet, P.~Van Mechelen, A.~Van~Spilbeeck\\

\textbf{Centro~Brasileiro~de~Pesquisas~Fisicas,~Rio~de~Janeiro,~Brazil}\\
G.A.~Alves, W.L.~Ald\'{a}~J\'{u}nior, C. Hensel\\

\textbf{Universidade~do~Estado~do~Rio~de~Janeiro,~Rio~de~Janeiro,~Brazil}\\
W.~Carvalho, J.~Chinellato, C.~De~Oliveira~Martins, D.~Matos~Figueiredo, C.~Mora~Herrera, H.~Nogima, W.L.~Prado~Da~Silva, E.J.~Tonelli~Manganote, A.~Vilela~Pereira\\

\textbf{Charles~University,~Prague,~Czech~Republic}\\
M.~Finger, M.~Finger~Jr., A.~Kveton, J.~Tomsa\\

\textbf{Institute~of~High~Energy~Physics~and~Informatization,~Tbilisi~State~University,~Tbilisi,~Georgia}\\
G.~Adamov, Z.~Tsamalaidze\footnotemark[1]\\

\textbf{Deutsches~Elektronen-Synchrotron,~Hamburg,~Germany}\\
U.~Behrens, K.~Borras, A.~Campbell, F.~Costanza, P.~Gunnellini, A.~Lobanov, I.-A.~Melzer-Pellmann, C.~Muhl, B.~Roland, M.~Sahin, P.~Saxena\\

\textbf{Indian~Institute~of~Science~Education~and~Research,~Pune,~India}\\
V.~Hegde, K.~Kothekar, S.~Pandey, S.~Sharma\\

\textbf{Panjab~University,~Chandigarh,~India}\\
S.B.~Beri, B. Bhawandeep, R.~Chawla, A.~Kalsi, A.~Kaur, M.~Kaur, G.~Walia\\

\textbf{Saha~Institute~of~Nuclear~Physics,~Kolkata,~India}\\
S.~Bhattacharya, S.~Ghosh, S.~Nandan, A.~Purohit, M.~Sharan\\

\textbf{Tata~Institute~of~Fundamental~Research-B,~Mumbai,~India}\\
S.~Banerjee, S.~Bhattacharya, S.~Chatterjee, P.~Das, M.~Guchait, S.~Jain, S.~Kumar, M.~Maity, G.~Majumder, K.~Mazumdar, M.~Patil, T.~Sarkar\\

\textbf{Vilnius~University,~Vilnius,~Lithuania}\\
A.~Juodagalvis\\

\textbf{Joint~Institute~for~Nuclear~Research,~Dubna,~Russia}\\
S.~Afanasiev, P.~Bunin, Y.~Ershov, I.~Golutvin, A.~Malakhov, P.~Moisenz{\def\thefootnote{\fnsymbol{footnote}}\footnotemark[2]}, V.~Smirnov, A.~Zarubin\\

\textbf{National~Research~Nuclear~University~Moscow~Engineering~Physics~Institute,~Moscow,~Russia}\\
M.~Chadeeva, R.~Chistov, M.~Danilov, E.~Popova, V.~Rusinov\\

\textbf{Institute~for~Nuclear~Research,~Moscow,~Russia}\\
Yu.~Andreev, A.~Dermenev, A.~Karneyeu, N.~Krasnikov, D.~Tlisov, A.~Toropin
% ...
\\

\textbf{University~of~Virginia,~Charlottesville,~USA}\\
S.~Goadhouse, R.~Hirosky, Y.~Wang\\

{\def\thefootnote{\fnsymbol{footnote}}\footnotetext[2]{Deceased}}
\footnotetext[1]{Also at Joint~Institute~for~Nuclear~Research,~Dubna,~Russia}
\footnotetext[2]{Also at California~Institute~of~Technology,~Pasadena,~USA}
\footnotetext[3]{Also at The~University~of~Alabama,~Tuscaloosa,~USA}
\footnotetext[4]{Also at Gaziosmanpasa~University,~Tokat,~Turkey}
\footnotetext[5]{Also at Adiyaman~University,~Adiyaman,~Turkey}
\footnotetext[6]{Also at Piri~Reis~University,~Istanbul,~Turkey}
\footnotetext[7]{Also at Necmettin~Erbakan~University,~Konya,~Turkey}
\footnotetext[8]{Also at Marmara~University,~Istanbul,~Turkey}
\footnotetext[9]{Also at Kafkas~University,~Kars,~Turkey}
\footnotetext[10]{Also at Mimar~Sinan~University,~Istanbul,~Turkey}
\footnotetext[11]{Also at Istanbul~University,~Istanbul,~Turkey}
\footnotetext[12]{Also at Istanbul~Bilgi~University,~Istanbul,~Turkey}
\footnotetext[13]{Also at Yildiz~Technical~University,~Istanbul,~Turkey}
\footnotetext[14]{Also at Hacettepe~University,~Ankara,~Turkey}
\footnotetext[15]{Also at Beykent University,~Istanbul,~Turkey}
\footnotetext[16]{Also at Erzincan~University,~Erzincan,~Turkey}
\footnotetext[17]{Now at NOAA,~National~Oceanic~and~Atmospheric~Administration,~USA}
\footnotetext[18]{Now at University~of~California,~Davis,~Davis,~USA}
\footnotetext[19]{Also at Institute~for~Nuclear~Research,~Moscow,~Russia}
\footnotetext[20]{Now at INFN~Sezione~di~Milano-Bicocca,~Milano,~Italy}

\end{document}